# How Large is the Elephant in the Density Functional Theory Room?

Frank Jensen

*Department of Chemistry, Aarhus University, Langelandsgade 140, DK-8000 Aarhus, Denmark*

A recent paper compares density functional theory results for atomization energies and dipole moments using a multi-wavelet based method with traditional Gaussian basis set results, and concludes that Gaussian basis sets are problematic for achieving high accuracy. We show that by a proper choice of Gaussian basis sets they are capable of achieving essentially the same accuracy as the multi-wavelet approach, and identify a couple of possible problems in the multi-wavelet calculations.

In a recent paper, Jensen et al. have reported multi-wavelet results using the MRChem program for Density Functional Theory (DFT) total and atomization energies (AE) as well as dipole moments for a benchmark set of data with 211 molecules.[1] They state that their method is capable of achieving micro-Hartree accuracy and compare their results to those obtained by numerical orbital and Gaussian-type orbitals methods. One of their conclusions was that Gaussian basis sets have difficulties achieving accuracies for atomization energies better than ~0.1 kcal/mol on average, and displaying outliers in the ~2 kcal/mol region even with very large basis sets. The purpose of the present work is to illustrate that Gaussian basis sets are in fact capable of achieving accuracies of ~0.01 kcal/mol for atomization energies, and point out that the reported MRChem results in some cases may be less accurate than stated.

The Jensen et al. paper used the aug-cc-pVXZ basis sets[2] for probing the performance of Gaussian type basis sets for the PBE and PBE0 functionals.[3,4] These basis sets, however, have been developed for correlated wave function methods, and are not optimum for DFT methods, displaying a relatively slow basis set convergence behavior.[5] The pc-n basis sets,[6] on the other hand, have been designed explicitly for DFT methods, and the most recent versions denoted pcseg-n have been defined for all atoms up to Kr and up to pentuble zeta quality.[7] These basis sets employ a segmented contraction scheme in order to improve the computational efficiency, but for assessing the contraction error, the corresponding uncontracted pc-n basis sets, denoted upc-n, can be employed.

We were initially unable to reproduce the reported total energies using standard Gaussian basis sets, but as shown in Table 1 for $BCl_3$, this reflects that Jensen et al. have used a grid for calculating the exchange-correlation part of the DFT energy that is not saturated in terms of grid points.[1] While the reported total energies with the standard basis sets thus may have errors in the ~0.1 milli-Hartree range due to grid errors, this fortunately cancels almost exactly when calculating atomization energies. We have in the present work employed grid sizes that provide total energies stable to at least 1 micro-Hartree.[8]

Table 1. Total (Hartree) and atomization (kcal/mol) energies for the $BCl_3$ molecule using the PBE functional and the 6-311++G(3df,3pd) basis set with different grid sizes for calculating the exchange-correlation energy term.

| Grid | | $E_{tot}$ | AE |
|---|---|---|---|
| NWChem | NWChem | -1405.025674 | 338.332 |
| Radial | Angular | | |
| 35 | 110 | -1405.026415 | 338.099 |
| 75 | 302 | -1405.025526 | 338.329 |
| 99 | 590 | -1405.025532 | 338.332 |
| 250 | 974 | -1405.025534 | 338.333 |
| 900 | 974 | -1405.025534 | 338.333 |

NWChem: values reported in ref. [1] using the NWChem program.

The mean and maximum absolute deviations (MAD and MaxAD) over the benchmark data set relative to the reported MRChem-T4 results, with the exception of $CH_3S$ which is discussed below, are shown in Table 2. At the double zeta level (cc-pVDZ, pcseg-1), there is little difference in the performance of the different basis sets. At the triple zeta level, the pcseg-2 basis set performs better than cc-pVTZ, and this difference accelerates at the quadruple zeta level (cc-pVQZ and pcseg-3), especially for the MaxAD values. At the pentuble zeta level, the difference between the pcseg-4 and upc-4 results shows that the contraction error now is the limiting factor for the pcseg-4 basis set.

Employing a square-root exponential extrapolation[9] of the pc-2,3,4 results provides the results denoted with Xpol, where the mean absolute deviation relative to the MRChem-T4 results is ~0.01 kcal/mol, with maximum deviations ~0.08 kcal/mol. Augmenting the basis sets with diffuse functions provide a small improvement, which diminished as the underlying basis set becomes more complete. There is very little difference in the performance for the two employed functionals, which suggests that these levels of accuracies should be valid for DFT methods in general.

Table 2. Mean and maximum absolute deviations (MAD and MaxAD) of atomization energies (kcal/mol) relative to the MRChem-T4 results over the benchmark set of data without the $CH_3S$ results.

|  | DFT | PBE | | | | | PBE0 | | | | |
| --- | --- | --- | --- | --- | --- | --- | --- | --- | --- | --- | --- |
|  | Basis/Level[a] | D | T | Q | 5 | Xpol | D | T | Q | 5 | Xpol |
| MAD | cc-pVXZ | 8.141 | 1.508 | 0.871 | 0.179 |  | 9.426 | 1.807 | 0.691 | 0.167 |  |
|  | aug-cc-pVXZ | 8.211 | 1.838 | 0.625 |  |  | 8.248 | 1.831 | 0.582 |  |  |
|  | pcseg-n | 8.119 | 1.103 | 0.158 | 0.107 | 0.106 | 8.234 | 1.073 | 0.096 | 0.058 | 0.067 |
|  | aug-pcseg-n | 5.735 | 0.551 | 0.107 | 0.097 | 0.100 | 4.854 | 0.461 | 0.073 | 0.053 | 0.063 |
|  | upc-n | 5.877 | 1.208 | 0.082 | 0.020 | 0.016 | 6.548 | 1.245 | 0.073 | 0.015 | 0.011 |
|  | aug-upc-n | 4.115 | 0.576 | 0.056 | 0.016 | 0.013 | 3.846 | 0.565 | 0.048 | 0.011 | 0.009 |
| MaxAD |  |  |  |  |  |  |  |  |  |  |  |
|  | cc-pVXZ | 48.72 | 13.96 | 7.74 | 1.79 |  | 60.44 | 16.67 | 8.97 | 2.07 |  |
|  | aug-cc-pVXZ | 54.75 | 20.94 | 10.78 |  |  | 55.13 | 19.82 | 10.18 |  |  |
|  | pcseg-n | 58.40 | 9.79 | 0.61 | 0.46 | 0.47 | 67.01 | 10.07 | 0.50 | 0.23 | 0.25 |
|  | aug-pcseg-n | 51.76 | 7.04 | 0.39 | 0.44 | 0.45 | 50.37 | 6.45 | 0.37 | 0.23 | 0.23 |
|  | upc-n | 50.90 | 9.90 | 0.54 | 0.09 | 0.08 | 57.69 | 10.47 | 0.50 | 0.09 | 0.06 |
|  | aug-upc-n | 47.92 | 6.71 | 0.34 | 0.09 | 0.08 | 47.28 | 6.25 | 0.37 | 0.10 | 0.11 |

[a]: The D/T/Q/5 notation for the cc-pVXZ basis sets corresponds to the pc-1,2,3,4 notation, respectively, in terms of basis set quality. Xpol indicates a square-root exponential extrapolation of the pc-2,3,4 results.[9]

In the analysis of the results for the benchmark set of data, we noted that $CH_3S$ had the largest discrepancy between the MRChem-T4 and Xpol results, and this is analyzed in more detail in Table 3.

Table 3 shows the total and atomization energies for the $CH_3SH$ and $CH_3S$ molecules at the MRChem-T1,2,3,4 levels and using the uncontracted versions of the pc-n and aug-pc-n basis sets. The $CH_3SH$ system behaves as expected, with the MRChem-T4 and basis set extrapolated atomization energies agreeing to within ~0.01 kcal/mol and to within ~0.1 milli-Hartree for the total energy with both the PBE and PBE0 functionals. The $CH_3S$ system, on the other hand, display differences in the atomization energy of ~0.20 kcal/mol for both the PBE and PBE0 functionals. The basis set calculated total energies with the largest basis set are furthermore lower than the MRChem-T1,2,3,4 results, and this points to possible problems in at least some of these calculations.

Table 3. Total (Hartree) and atomization (kcal/mol) energy convergence for the $CH_3SH$ and $CH_3S$ molecules. T1,2,3,4 results are obtained with the MRChem program and taken from ref. [1], the other results are obtained using the uncontracted pc-n and aug-pc-n basis sets, where uxpol and auxpol indicated results obtained by a square-root exponential extrapolation of the pc-2,3,4 results.[9]

|  | $CH_3SH$ | | | | $CH_3S$ | | | |
|  | PBE | | PBE0 | | PBE | | PBE0 | |
|  | $E_{tot}$ | AE | $E_{tot}$ | AE | $E_{tot}$ | AE | $E_{tot}$ | AE |
| T1 | -438.51282 | 477.228 | -438.55250 | 473.131 | -437.87170 | 388.730 | -437.91072 | 385.043 |
| T2 | -438.51282 | 477.719 | -438.55260 | 472.688 | -437.87156 | 389.074 | -437.91062 | 384.423 |
| T3 | -438.51282 | 477.735 | -438.55260 | 472.709 | -437.87154 | 389.076 | -437.91059 | 384.425 |
| T4 | -438.51282 | 477.737 | -438.55261 | 472.713 | -437.87154 | 389.077 | -437.91059 | 384.426 |
|  |  |  |  |  |  |  |  |  |
| upc-1 | -438.46075 | 472.093 | -438.50043 | 466.805 | -437.82331 | 385.313 | -437.86215 | 380.361 |
| upc-2 | -438.50586 | 476.618 | -438.54574 | 471.662 | -437.86544 | 388.459 | -437.90447 | 383.800 |
| upc-3 | -438.51234 | 477.742 | -438.55216 | 472.719 | -437.87137 | 389.270 | -437.91040 | 384.588 |
| upc-4 | -438.51270 | 477.757 | -438.55251 | 472.730 | -437.87173 | 389.289 | -437.91076 | 384.610 |
| uxpol | -438.51274 | 477.752 | -438.55254 | 472.725 | -437.87177 | 389.287 | -437.91079 | 384.608 |
|  |  |  |  |  |  |  |  |  |
| aupc-1 | -438.46584 | 473.542 | -438.50544 | 468.811 | -437.82759 | 386.318 | -437.86642 | 381.923 |
| aupc-2 | -438.50718 | 477.169 | -438.54696 | 472.180 | -437.86661 | 388.916 | -437.90561 | 384.271 |
| aupc-3 | -438.51242 | 477.742 | -438.55223 | 472.709 | -437.87147 | 389.282 | -437.91050 | 384.597 |
| aupc-4 | -438.51271 | 477.746 | -438.55252 | 472.719 | -437.87175 | 389.286 | -437.91078 | 384.607 |
| auxpol | -438.51274 | 477.743 | -438.55254 | 472.717 | -437.87178 | 389.284 | -437.91081 | 384.606 |

Jensen et al. also examined the dipole moment as a property that is not directly related to the energy, and indicated that basis set methods have difficulties attaining accuracies better than ~0.01 Debye.[1] Table 4 shows a comparison between the MRChem and basis set results, analogous to the energetic results in Table 2.

Table 4. Mean and maximum absolute deviations (MAD and MaxAD) of molecular dipole moments (Debye) relative to the MRChem-T4 results over the benchmark set of data without the $CH_3S$ and $CH_3NH_2$ results.

|  | DFT | PBE | | | | PBE0 | | | |
|  | Basis/Level[a] | D | T | Q | 5 | D | T | Q | 5 |
| MAD | cc-pVXZ | 0.1355 | 0.0624 | 0.0272 | 0.0126 | 0.1098 | 0.0495 | 0.0213 | 0.0095 |
|       | aug-cc-pVXZ  | 0.0190 | 0.0084 | 0.0046 |        | 0.0206 | 0.0089 | 0.0051 |        |
|       | pcseg-n      | 0.1123 | 0.0467 | 0.0101 | 0.0027 | 0.1163 | 0.0468 | 0.0086 | 0.0019 |
|       | aug-pcseg-n  | 0.0189 | 0.0060 | 0.0008 | 0.0004 | 0.0213 | 0.0066 | 0.0009 | 0.0005 |
|       | upc-n        | 0.1162 | 0.0525 | 0.0102 | 0.0021 | 0.1183 | 0.0519 | 0.0089 | 0.0015 |
|       | aug-upc-n    | 0.0161 | 0.0054 | 0.0005 | 0.0007 | 0.0186 | 0.0061 | 0.0005 | 0.0002 |
| MaxAD |              |        |        |        |        |        |        |        |        |
|       | cc-pVXZ      | 1.212  | 0.648  | 0.326  | 0.156  | 1.163  | 0.619  | 0.309  | 0.135  |
|       | aug-cc-pVXZ  | 0.218  | 0.119  | 0.091  |        | 0.234  | 0.123  | 0.095  |        |
|       | pcseg-n      | 0.870  | 0.293  | 0.069  | 0.038  | 1.016  | 0.269  | 0.062  | 0.021  |
|       | aug-pcseg-n  | 0.155  | 0.072  | 0.011  | 0.002  | 0.171  | 0.077  | 0.010  | 0.005  |
|       | upc-n        | 0.894  | 0.326  | 0.057  | 0.020  | 0.951  | 0.345  | 0.064  | 0.016  |
|       | aug-upc-n    | 0.141  | 0.069  | 0.007  | 0.009  | 0.153  | 0.076  | 0.008  | 0.003  |

[a]: The D/T/Q/5 notation for the cc-pVXZ basis sets corresponds to the pc-1,2,3,4 notation, respectively, in terms of basis set quality.

It is well-known that the basis set convergence for electric moments is greatly accelerated by employing basis sets augmented with diffuse functions,[10] and this is clearly displayed by the results in Table 4. The aug-pc-n type basis sets again display a better and more uniform convergence than the aug-cc-pVXZ type basis sets. Besides the already mentioned problem with the $CH_3S$ system, we also observed a discrepancy for the $CH_3NH_2$ system, which is analyzed in more detail in Table 5.

Table 5. Molecular dipole moment (Debye) convergence for the $CH_3CH_2NH_2$ and $CH_3NH_2$ compounds. T1,2,3,4 results are obtained with the MRChem program and taken from ref.[1], the other results are obtained using the uncontracted aug-pc-n basis sets.

|        | $CH_3CH_2NH_2$ |        | $CH_3NH_2$ |        |
|--------|--------|--------|--------|--------|
|        | PBE    | PBE0   | PBE    | PBE0   |
| T1     | 1.2960 | 1.3509 | 1.3043 | 1.2748 |
| T2     | 1.2989 | 1.3559 | 1.3061 | 1.2766 |
| T3     | 1.3008 | 1.3569 | 1.3067 | 1.2775 |
| T4     | 1.3011 | 1.3570 | 1.3069 | 1.2775 |
|        |        |        |        |        |
| aupc-1 | 1.3073 | 1.3648 | 1.2416 | 1.2822 |
| aupc-2 | 1.3046 | 1.3603 | 1.2420 | 1.2814 |
| aupc-3 | 1.3013 | 1.3571 | 1.2380 | 1.2776 |
| aupc-4 | 1.3011 | 1.3571 | 1.2377 | 1.2776 |

The PBE and PBE0 results for $CH_3CH_2NH_2$ display excellent agreement between the MRChem and basis set results, but the MRChem PBE dipole moment for $CH_3NH_2$ deviates by 0.069 Debye relative to the aug-upc-4 results, and this indicates a possible error. We also note that several of the dipole moments which must be identical zero due to the molecular symmetry, are reported to have non-zero MRChem values, with the largest deviation being $BO_2$ with a value of 0.0015 Debye.[1] These deviations are comparable to the differences between the MRChem and aug-upc-4 results in Table 4.

Jensen et al. states that Gaussian basis sets cannot be systematically enlarged to achieve $L^2$ completeness, but the results for the (aug-)pc-n basis sets suggest that the basis set error can be reduced well below chemical accuracy. There are no fundamental limitations in designing higher level pc-n basis sets, and we have in special cases employed pc-5 and pc-6 type basis sets for even higher accuracy.[11, 12] For practical purposes, however, there is little gained by employing basis sets larger than (aug-)pc-3, as method errors then dominates the results.

The chemistry and physics communities have traditionally employed different methods and technologies for calculating the electronic structure of molecules and periodic systems. The ability to calculate the same quantities by different methodologies allows an independent validation of the results and identification of possible problems. The paper by Jensen et al. is an important contribution to this field,[1] but the present work shows that care must be taken for generating reference results of very high accuracy as for example ~0.01 kcal/mol for atomization energies and ~0.0001 Debye for dipole moments. The present work also shows that the use of basis sets optimized for the employed Hamiltonian is essential for achieving the full accuracy of the methodology.[13, 14]

**ACKNOWLEDGMENT**

This work was supported in part by the Danish Natural Science Research Council by Grant No. 4181-00030B.